\documentclass[%
reprint,                     
amsmath,amssymb,
aps,
]{revtex4-2}

\usepackage{graphicx}
\usepackage{dcolumn}
\usepackage{bm}
\usepackage{hyperref}

\usepackage{xcolor}
\usepackage{soul}

\usepackage{comment}
\begin{document}
	
	\preprint{APS/123-QED}
	
	\title{{Bistability-assisted Mechanical Squeezing and Entanglement}}%
	
	\author{Souvik Agasti}
	\email{souvik.agasti@uhasselt.be}
	\affiliation{
		IMOMEC division, IMEC, Wetenschapspark 1, B-3590 Diepenbeek, Belgium
	}%
	\affiliation{
		Institute for Materials Research (IMO), Hasselt University,	Wetenschapspark 1, B-3590 Diepenbeek, Belgium
	}%
	
	\author{P. Djorw\'e}
	\email{djorwepp@gmail.com}
	\affiliation{Department of Physics, Faculty of Science, 
		University of Ngaoundere, P.O. Box 454, Ngaoundere, Cameroon}

	
	\begin{abstract}
		
		We propose a scheme to squeeze mechanical motion and to entangle optical field with mechanical motion in an optomechanical system containing a parametric amplification. The scheme is based on optical bistability which emerges in the system for a strong enough driving field. By considering the steady state's lower branch of the bistability, the system shows weak entanglement and almost no mechanical squeezing. When the steady state is on the upper branch of the bistable shape, both squeezing and entanglement are greatly enhanced. Specifically, the entanglement shows three degrees of magnitude enhancement. However, this giant entanglement is fragile against decoherence and thermal fluctuation.  Regarding the mechanical squeezing, it reaches the standard quantum limit (SQL) in the upper branch of the bistability. Our proposal provides a way to improve quantum effects in optomechanical systems by taking advantage of nonlinearities. This scheme can be realized in similar systems such as superconducting microwave, and hybrid optomechanical systems.
		
	\end{abstract}
	
	\maketitle
	

	\section{Introduction}\label{Introduction}
	
	The precision measurement of the motion of a harmonic oscillator can be minimized to ground-state quantum zero point fluctuation, which is limited by the Heisenberg uncertainty principle ($\Delta X_1 \Delta X_2 \geq 1 $, where $ X_1 $ are the $ X_2 $ are its two motional quadratures). One can, in principle, minimize the fluctuation in one quadrature with the cost of an equivalent increment on the other one.  This process is known as quantum squeezing, and it was primarily generated on the squeezed laser in the early 80’s \cite{squeezed_light, squeezed_light_4_wave_mix}.  Since that, quantum squeezing has been used in various experiments including mechanical motion and optical field in cavity optomechanical systems for instance \cite{philippe_squeezing, clark_sideband_sqz, squeeze_photon_optomechanics_reservoir_engineering}.  Squeezed states are interesting for a range of applications including force-sensing \cite{Vitali_suggestion_BAE, Force_sense_Courty}, exceptional point sensors \cite{Philippe_Exceptional_Point, Exceptional_Point_sensor_1, Exceptional_Point_sensor_2, Philippe_Exceptional_Point_2}  and, overall interferometric detectors such as those in LIGO \cite{Advanced_LIGO} and VIRGO \cite{Advanced_Virgo} experiments. 
	
	A lot of effort has been made over decades to squeeze mechanical oscillators below standard quantum limit (SQL), for instance by using frequency modulation \cite{squeeze_photon_optomechanics_frequency_modulation}, two-tone driving \cite{clark_sideband_sqz} and reservoir engineering \cite{squeeze_photon_optomechanics_reservoir_engineering, reservoir_engineering}. Besides, it is also seen that periodical modulation of the amplitude of external drive can induce a high degree of mechanical squeezing along with an optomechanical entanglement in a cavity optomechanical system \cite{Optomechanics_periodic_modulation}. The ground-state cooling of the mechanical mode, usually required to achieve this goal, has been obtained through sideband cooling. Such schemes can be considered as an example of reservoir engineering \cite{reservoir_engineering} where the cavity acts as a reservoir whose force noise is squeezed. Squeezed states of light are typically generated using Kerr nonlinear systems through parametric down-conversion process \cite{Quantum_optics_Walls_Milburn}. In this direction, the dynamical Casimir effect (DCE), i.e. quick movement of the mirror can generate photon pairs from a quantum vacuum, has been implemented in cavity optomechanical systems \cite{DCE_optomechanics_1, DCE_optomechanics_2}, where a parametric conversion of mechanical energy to optical photons has been observed. However, to observe the DCE, the mirror velocity is required to be close to the velocity of light, which is very difficult to achieve in practice. To bypass this problem, a few alternative schemes have been  proposed, e.g. modulating the ultrastrong light-matter coupling in cavity quantum electrodynamics (QED) or driving an optical parametric oscillator. In the present work, we explore both of them to reach the limit to obtain DCE.  
	
	Our benchmark system is an optomechanical cavity containing a parametric amplifier, which is used to induce optical bistability once the driving field is strong enough. By applying a squeezing operator to diagonalize the Hamiltonian, a DCE is induced in an observable range \cite{DCE_optomechanics_Nori}. With experimentally feasible parameters, both the DCE and the bistability are used to enhance optical steady states variables in the upper branch of the bistable shape. By using these enhanced optical steady states, we observed a giant enhancement of mechanical squeezing together with a greatly enhancement of entanglement between optical and mechanical modes. The generated entanglement is three order of magnitude greater than the one generated by the lower branch optical steady states. This work provides an alternative way of generating nonclassical states by exploring nonlinear effects in optomechanical systems. This scheme can be extended to similar setup such as superconducting microwaves, and hybrid optomechanical systems.
	
	The rest of this paper is organized as follows. In section II, we describe the model and the dynamical equations. The bistability-assisted squeezing and entanglement, together with the effect of the squeezing parameter, are described in section III, while section IV concludes the work.

	\section{MODEL: Dynamical Casimir Effect}
	
\begin{figure*}[t!]
		\includegraphics[width= 0.9 \linewidth]{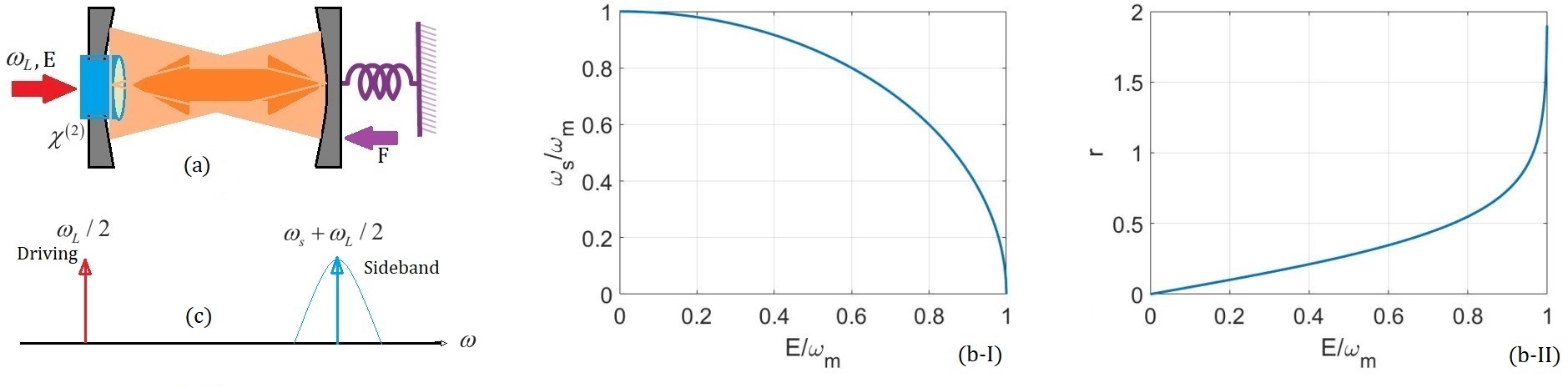}
		\caption{(a) Bloc diagram of parametrically driven cavity optomechanical system.	(b-I) Effective cavity frequency in squeezed frame and (b-II) squeezing factors with the variation of parametric drive, for $\Delta_s = \omega_m$. (c) illustration in frequency-domain of mechanically induced two-photon hyper-Raman scattering. The left arrow is the two-photon driving ($\omega_L/2$), and the right arrow is the squeezing-induced anti-Stokes cavity-sideband ($\omega_s + \omega_L/2$).}\label{Cashimir_optomechanics}
	\end{figure*}

	We consider an optomechanical system detuned by a two-photon drive of frequency $\omega_L$ and amplitude $E$, which eventually squeezes the cavity mode. The system is sketched in Fig. \ref{Cashimir_optomechanics}(a) where the driving induces the parametric down-conversion of mechanical phonons to generate correlated cavity-photon pairs, which corresponds to the Dynamical Casimir Effect (DCE). The cavity detuning $\Delta_s$ depends on the cavity frequency $(\omega_c)$ and driving frequency $(\omega_L)$ as, $\Delta_s = \omega_c -\omega_L/2 $), which can be useful to tune the parametric phonon-photon coupling into resonance. We also consider that the mechanical mode is driven, which leads to the system's Hamiltonian,
	\begin{equation}\label{full_hamiltonian}
		H = H_{CD} + H_{OM} + H_{MD},
	\end{equation}
	with,
	\begin{subequations}
		\begin{align}
			H_{CD} &= \Delta_s a_s^\dagger a_s + \frac{E}{2}(a_s^2+ {a_s^\dagger}^2) \label{cav_detune_Hamil},\\
			H_{OM} &= \omega_m b^\dagger b - g_{0} a^\dagger_s a_s(b+b^\dagger) \label{optomec_Hamil},\\
			H_{MD} &= \frac{F}{2} [e^{i\omega_d t} b + e^{-i\omega_d t} b^\dagger],
		\end{align}
	\end{subequations}
	where $a_s(a_s^\dagger)$ and $b(b^\dagger)$ are the annihilation (creation) operators of the cavity and mechanical modes, respectively. The single-phonon mechanical drive has the frequency $\omega_d$ and amplitude $F$. Under the influence of the parametric drive, the bare cavity mode is squeezed by a squeezing parameter,
	\begin{equation} \label{sq_parameter}
		r = \frac{1}{4} \ln \left( \frac{\Delta_s + E}{\Delta_s -E} \right),
	\end{equation}
	which transforms the cavity field operators to a squeezed mode, through the Bogoliubov transformation,
	\begin{equation}\label{Bogoliubov_transformation}
		a =  a_s\cosh r  + a_s^\dagger\sinh r.
	\end{equation}
	This Bogoliubov transformation diagonalizes the cavity detuning Hamiltonian $H_{CD}$ in Eq. \eqref{cav_detune_Hamil} to $\omega_s a^\dagger a$, where $ \omega_s = \sqrt{\Delta_s^2 - E^2} $. The effective cavity frequency $\omega_s$ and the squeezing parameter $(r)$ are depicted versus the parametric drive in Fig. \ref{Cashimir_optomechanics}(b-I,II). It can be seen that $\omega_s$ decreases over the parametric driving $E$, while the squeezing parameter $r$ increases. We have considered $\Delta_s = \omega_m$ for this analysis, which is a usual condition to witness optomechanical interaction in the red optomechanical sideband. Similar methods have been used for enhancing light-matter interactions in cavity optomechanics \cite{light_matter_optomechanics} and cavity QED \cite{light_matter_QED}. After the diagonalization, the Hamiltonian $H_{OM}$ yields,	
	\begin{equation}
		H_{OM} = \omega_m b^\dagger b + [ - g_{OM} a^\dagger a + G_0 (a^2+ {a^\dagger}^2) ] (b+b^\dagger)	
	\end{equation}
	where $g_{OM} = g_0 \cosh (2r) $ is an effective single-photon optomechanical coupling,  and $G_0 = g_0 \sinh (2r)/2 $ is a coupling associated with the DCE. Under the rotating wave approximation and by moving to the frame of mechanical drive, the Hamiltonian in Eq. \ref{full_hamiltonian} leads to an effective Hamiltonian,	
	\begin{equation}\label{effective_hamil}
		{H}_{eff} = \Delta a^\dagger a +\Omega_M b^\dagger b +G_0 \left( b^\dagger a^2+ b {a^\dagger}^2 \right) +\frac{1}{2} (F^* b+ F b^\dagger) 
	\end{equation}
	where $\Delta= \omega_s-\omega_d/2$ and $\Omega_M = \omega_m - \omega_d $ are the effective cavity and mechanical resonance frequency, respectively. The DCE in the squeezed frame can be understood as mechanically induced	two-photon hyper-Raman scattering in the laboratory frame. Fig. \ref{Cashimir_optomechanics}(c) shows that the hyper-Raman scattering is an anti-Stokes process. The squeezing gives rise to an anti-Stokes cavity-sideband at frequency $\omega_s+\omega_L/2$. The driving at frequency $\omega_L$ produces photon pairs at frequency $\omega_L/2$. The anti-Stokes scattering of driving photon pairs is anticipated to happen by absorbing mechanical phonons which are at frequency $\omega_m \approx 2\omega_s$. 
	
	\begin{figure*}[t!]
		\includegraphics[width=0.6\linewidth]{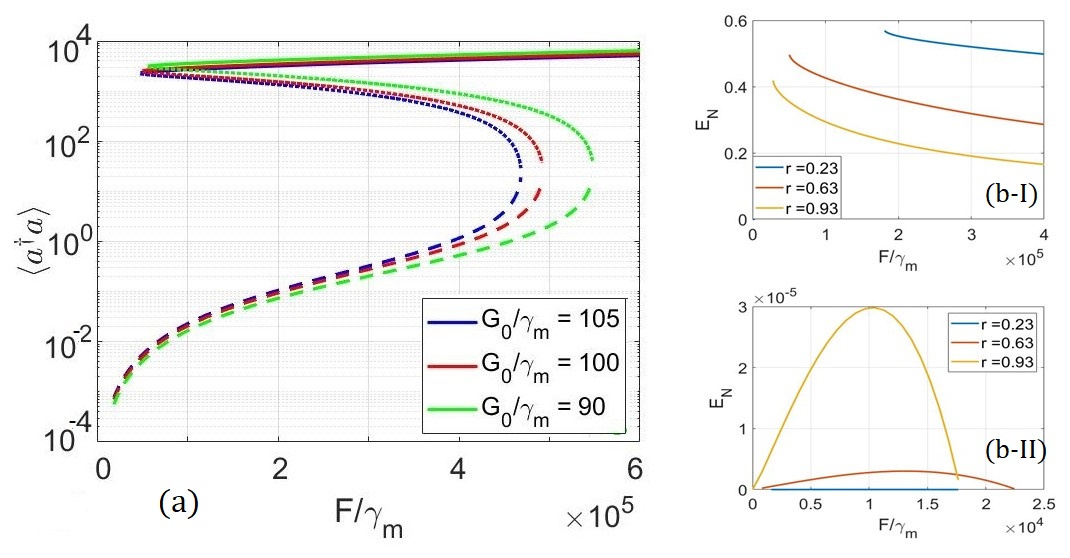}
		\caption{ (a) Bistability in the cavity population satisfying the resonant condition $\Delta = \Omega_M/2$. Solid, dotted, and dashed lines correspond to upper, unstable, and lower branches, respectively. The other parameters chosen here are $ \kappa= 500 \gamma_m, \Omega_M = 10^{4}\gamma_m $ and  $n_m = 0$. (b) Entanglement in (I) upper and (II) lower branches where the system is operated in the red detuning regime, i.e. $\Delta_s = \omega_m$.  $g_0= 1.15 \times 10^{2} \gamma_m$ and  the squeezing parameter is monitored by tuning cavity drive $E = 0.43 \omega_m, 0.87 \omega_m, 0.95 \omega_m$ which also gives $\omega_s = 0.9 \omega_m, 0.5 \omega_m, 0.3 \omega_m$, respectively, for blue, red and yellow lines.}
		\label{cav_population_bistable}
	\end{figure*}
	
	\begin{figure*}[t!]
		\includegraphics[width=1\linewidth]{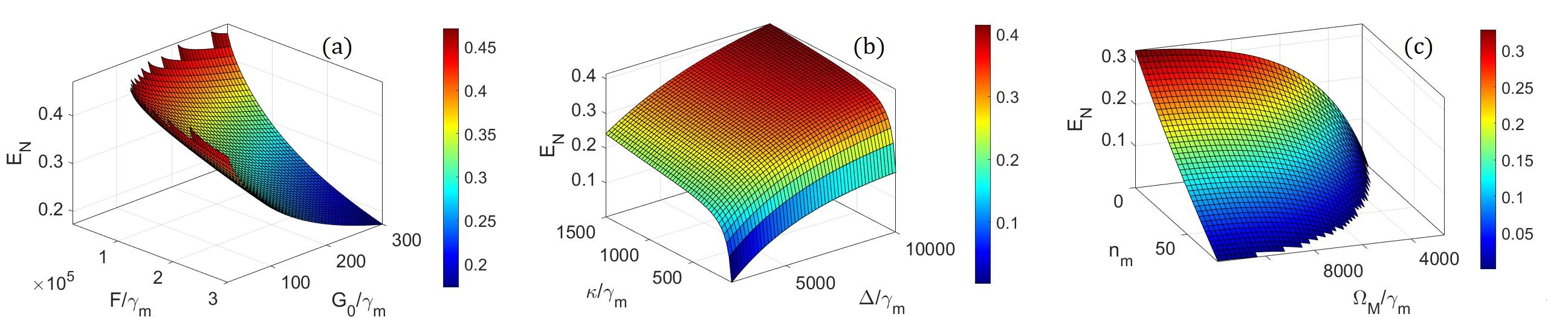}
		\caption{ Entanglement between cavity and mechanics for system saturating upper branch, with the variation of (a) mechanical drive and optomechanical coupling, (b) cavity detuning and linewidth, and (c) mechanical oscillation frequency and reservoir thermal population. All other parameters remain the same with Fig. \ref{cav_population_bistable} }
		\label{Entanglement_upper_brunch}
	\end{figure*}
	
	\begin{figure*}[t!]
		\includegraphics[width=1\linewidth]{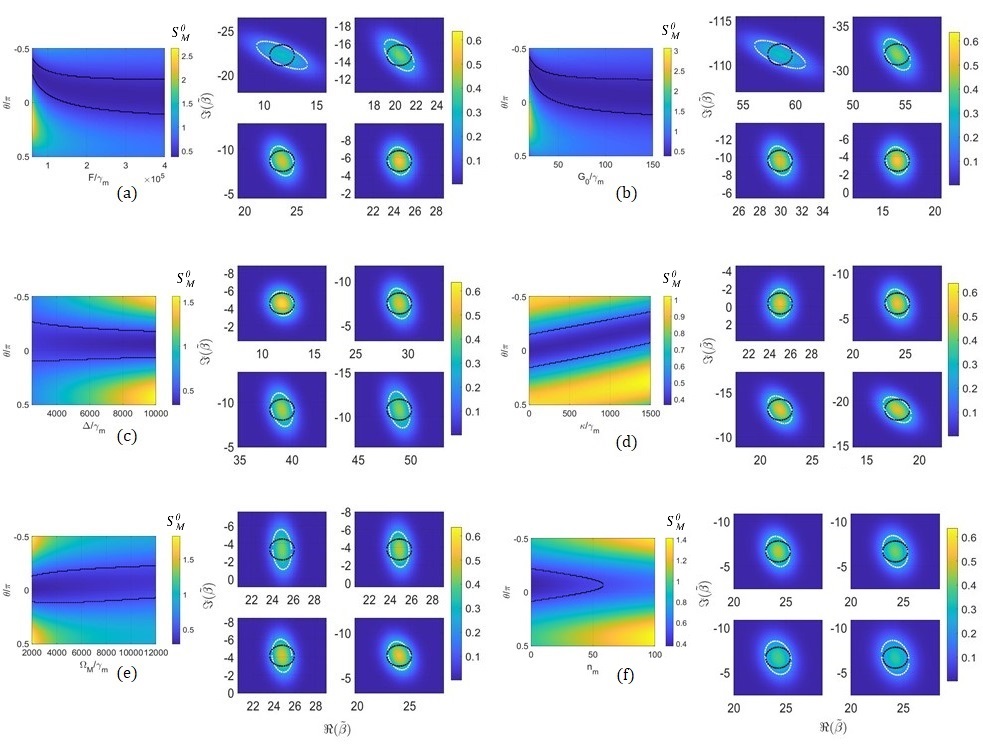}
		\caption{ Squeezing of mechanical motion in different quadratures and Wigner functions for different parameters, such as (a) strength of mechanical drive ($ F/\gamma_m = 6 \times 10^4; 10 \times 10^4; 20 \times 10^4; 40 \times 10^4 $), (b) DCE generated optomechanical coupling ( $ G_0/\gamma_m = 20; 40; 80; 150  $), (c) cavity detuning ( $ \Delta/\gamma_m = 2500,6000,8000, 10^{4}  $), (d) cavity linewidth ($ \kappa/\gamma_m = 50; 500; 1000; 1500  $), (e) frequency of mechanical oscillator ( $ \Omega_M/\gamma_m = 0.2\times 10^4; 0.3\times 10^4; 0.4\times 10^4; 1.2\times 10^4  $), and (f) thermal population of mechanical reservoir $(n_m = 25;50;75;100)$. All other parameters remain the same with Fig. \ref{cav_population_bistable}.  The black dashed lines represent the boundary of SQL in the plot of fluctuation of mechanical quadratures. White dashed lines in the plots of Wigner functions represent half of the maxima, and black dashed lines are for the corresponding coherent states, representing SQLs.}\label{wigfn_together}
	\end{figure*}
	
	\section{Bistability induces mechanical squeezing and entanglement}
	The dynamical equations of our system can be derived through Heisenberg-Langevin equations. From the effective Hamiltonian mentioned in Eq. \eqref{effective_hamil}, one gets
	\begin{subequations}\label{EOM_Heisenberg-Langevin}
		\begin{align}
			\dot{a} &= -(\kappa + i \Delta ) a -2 iG_0 a^\dagger b  +\sqrt{2\kappa} a_{in},  \\ 
			\dot{a}^\dagger &=  -(\kappa - i \Delta ) a^\dagger +2 iG_0 a b^\dagger +\sqrt{2\kappa} a^\dagger_{in},  \\
			\dot{b} &= -(\gamma_m + i \Omega_M ) b - iG_0 a^2 + \sqrt{2\gamma_m} b_{in} -i F/2, \\ 
			\dot{b}^\dagger &= -(\gamma_m - i \Omega_M ) b^\dagger + iG_0 { a^\dagger}^2 + \sqrt{2\gamma_m} b_{in}^\dagger  +i F^*/2 ,
		\end{align}
	\end{subequations}
	where  $\kappa$ ($\gamma_m$) is the optical (mechanical) dissipation rate, and $a_{in}$  ($b_{in}$) is the reservoir noise operator for the cavity (mechanics). Since the cavity oscillates at high frequency ($\omega_c$), the thermal noise effect on the cavity is negligible, $n_c(\omega_c)\delta(t-t') =\langle  a_{in}^\dagger(t)  a_{in}(t') \rangle \approx 0$. The contribution of thermal noise to the mechanics is quantized by $n_m (\omega_m)\delta(t-t') = \langle b_{in}^\dagger(t) b_{in}(t') \rangle$.  By splitting the cavity and mechanical fields into their steady-state and fluctuation ($a= \alpha + \delta a$ and $b = \beta +\delta b$), one can obtain steady states fields, by solving zeroth order of the EOMs given in Eq \eqref{EOM_Heisenberg-Langevin}, as
	
	\begin{subequations}
		\begin{align}
			\frac{d}{dt} \langle a^\dagger a \rangle &= -4 G_0 \Im[ \langle a^2 \rangle \langle b \rangle^*] -2\kappa \langle a^\dagger a \rangle, \\
			\frac{d}{dt} \langle a^2 \rangle &=	-2i \Delta  \langle a^2 \rangle - 2i G_0 (2  \langle a^\dagger a \rangle +1)\langle b \rangle - 2\kappa  \langle a^2 \rangle, \\
			\frac{d}{dt} \langle b \rangle &= -i \left( \Omega_M \langle b \rangle + G_0 \langle a^2 \rangle + \frac{F}{2} \right) - \gamma_m \langle b \rangle, 
		\end{align}
	\end{subequations}
	which leads to the steady-state mean field of the cavity and the mechanics,
	\begin{subequations}
		\begin{align}
			\langle a^2 \rangle_{ss} &= -\frac{F G_0 z}{2 \left(i \gamma_m  \Delta -\Delta  \Omega_M + G_0^2 z +\gamma_m \kappa+i \kappa \Omega_M \right)},\\
			\langle b\rangle_{ss} &= -\frac{ F ( ik - \Delta )}{2 \left(i \gamma_m  \Delta -\Delta \Omega_M+ G_0^2 z +\gamma_m  \kappa +i \kappa \Omega_M\right)},
		\end{align}
	\end{subequations}
	where  $z = (2 \langle a^\dagger a \rangle_{ss} +1) $ is obtained from the roots of the cubic equation,
	
	\begin{widetext}
		\begin{equation} \label{cav_pop}
			F^2 G_0^2 z - (z-1) \left(2 G_0^2 z (\gamma_m  \kappa -\Delta  \Omega_M)+ G_0^4 z^2+ (\gamma_m ^2+ \Omega_M^2 ) (\Delta ^2+ \kappa^2 )\right) = 0 .
		\end{equation}
	\end{widetext}
	
	To get an insight into the DCE, we determine the steady states for the realistic parameters used in
	\cite{DCE_optomechanics_Nori}. However, the cavity and mechanical drives used in that work are not strong enough to induce bistability which is of interest. Therefore, we have increased the strength of both drives,  which enhances the DCE in our proposal. Besides that, it is also required to satisfy the resonant cavity detuning condition $\omega_m = 2\omega_c$, therefore $\Delta \approx \Omega_m/2$, to ensure the parametric energy of the mechanical motion to be delivered to the electromagnetic field \cite{DCE_optomechanics_Nori}.
	The cubic equation given in Eq. \eqref{cav_pop} induces DCE-generated bistability of the population of the cavity mode depicted in Fig. \ref{cav_population_bistable}(a). In order to enhance mechanical squeezing and entanglement, we will use the optical steady states lying on the upper branch of this $S$-like the shape of this bistable behavior. 
	The DCE-generated bistability in optomechanical systems has also been reported before in \cite{DCE_optomechanics_Nori}. Also, such non-linearity generated bistability is often observed in Kerr nonlinear systems \cite{Kerr_Drummond_Walls, Agasti_kerr}, and the hysteresis has also been observed within the range of transition region \cite{hysteresis_kerr_experimental_josa, Agasti_kerr_physica_scripta}. 
	
	Both squeezing and entanglement are characterized by mean of the amplitude and phase quadratures of the cavity $(X= (\delta a+\delta a^\dagger)/\sqrt{2}, Y = -i(\delta a-\delta a^\dagger)/\sqrt{2} )$ and mechanics ($ Q= (\delta  b+\delta b^\dagger)/\sqrt{2}, P = -i(\delta b-\delta b^\dagger)/\sqrt{2}$ ).  These quadratures can be deduced from the first order of the field operators in Eq. \eqref{EOM_Heisenberg-Langevin} as,
	
	\begin{equation}\label{EOM_t}
		\dot{\mathbf{u}} = 
		\mathbf{A_M} \mathbf{u} + \mathbf{n}, 
	\end{equation}
	where $\rm{u}=[X, Y,Q,P]^T $ and $\rm{n}=[ \sqrt{2\kappa} X_{in}, \sqrt{2\kappa} Y_{in}, \sqrt{2\gamma} Q_{in}, \sqrt{2\gamma} P_{in}] ^T$ are the vectors of system and noise quadrature operators, respectively, with  $(X_{in}= (a_{in}+a_{in}^\dagger)/\sqrt{2}, Y_{in} = -i(a_{in}-a_{in}^\dagger)/\sqrt{2} )$ and ($ Q_{in} = ( b_{in}+b_{in}^\dagger)/\sqrt{2}, P_{in} = -i(b_{in}-b_{in}^\dagger)/\sqrt{2}$) the external noise operators of the cavity and mechanical quadratures. The stability of the system can be analyzed by determining eigenvalues of the matrix $\mathbf{A_M}$,	
	
	\begin{widetext}
		\begin{equation} \label{A_M_coeff}
			\mathbf{A_M} = \left(
			\begin{array}{cccc}
				-\kappa + 2 G_0 \Im(\beta) & \Delta -2 G_0 \Re(\beta) & -2 G_0 \Im(\alpha) &  2 G_0 \Re(\alpha) \\
				-\Delta  -2 G_0 \Re(\beta) & -\kappa - 2 G_0 \Im(\beta) &  - 2 G_0 \Re(\alpha) &  -2 G_0 \Im(\alpha) \\
				2 G_0 \Im(\alpha) &  2 G_0 \Re(\alpha) & - \gamma_m & \Omega _M \\
				- 2 G_0 \Re(\alpha) & 2 G_0 \Im(\alpha) & - \Omega_M & - \gamma_m \\
			\end{array}\right).
		\end{equation}
	\end{widetext}
	
	According to the Routh–Hurwitz criteria, the real part of the eigenvalues of $\mathbf{A_M}$ must be negative for the system to be stable. Following this criterion, we found that the lower and upper branches of the bistability shown in Fig. \ref{cav_population_bistable}(a) are stables, while the middle branch is unstable. This allows us to determine the steady-state correlation matrix between the position and momentum operators of the optomechanical system $ (\mathbf{V} = \langle \{ \mathbf{u}, \mathbf{u}^T\}\rangle/2 )$  by solving Lyapunov equation as,
	\begin{equation}
		\rm{A_M} \rm{V} + \rm{V} \rm{A_M}^T = - \rm{D}, 
	\end{equation}
	where $ \rm{D} = \langle \{ \rm{n}, \rm{n}^T\}\rangle/2 = \text{diag}[ \kappa, \kappa, (n_m + 1/2)2\gamma_m ,  (n_m + 1/2)2\gamma_m ]$.
	
	Furthermore, we study entanglement between mechanics and cavity modes when the cavity saturates to the upper branch. The steady-state entanglement between two systems is calculated employing logarithmic negativity, defined by,
	\begin{equation}
		E_N = \max[0, -\ln 2\eta^-],
	\end{equation}
	where
	\begin{equation}
		\eta^- = \sqrt{\frac{ 1}{2} \left( \Sigma (\mathbf{V}) - \sqrt{\Sigma (\mathbf{V})^2 - 4 \det (\mathbf{V})} \right)},
	\end{equation}
	with $ \Sigma(\mathbf{V} ) = \det \mathbf{V_C} + \det \mathbf{V_M} - 2 \det \mathbf{V_{CM} }$ and 
	\begin{equation}\label{define_Vmatt}
		\mathbf{V} = \begin{bmatrix}
			\mathbf{V_C} &  \mathbf{V_{CM}} \\
			\mathbf{V_{CM}}^T &  \mathbf{V_M}
		\end{bmatrix}. 
	\end{equation}
	It is interesting to mention that the elements of the $\rm{V's}$ matrix are given by $\rm{V_{ij}} = \frac{1}{2} \langle u_i u_j + u_j u_i \rangle$. From the aforementioned definition of the logarithmic negativity, a Gaussian state is entangled $(\rm{E_N} > 0)$ if and only if $\rm{\eta}^- < \frac{1}{2}$, which is equivalent to Simon's necessary and sufficient entanglement nonpositive partial trace criteria for Gaussian states \cite{Simon}. We study the cavity-mechanical entanglement for the upper Bistability branch in Fig \ref{cav_population_bistable}(b-I). We observe that, while increasing cavity drive (which also increases cavity squeezing), the entanglement decreases rapidly. However, the entanglement is moreover seen to increase in the lower branch in Fig. \ref{cav_population_bistable}(b-II), even though it overall remains weak. Therefore, we target on upper branch to generate optomechanical entanglement and to observe entanglement-assisted quantum phenomenon.

	Fig. \ref{Entanglement_upper_brunch} shows the logarithmic negativity plotted in $\rm{3D}-$space versus certain parameters. In Fig. \ref{Entanglement_upper_brunch}(a), the entanglement is shown versus both mechanical drive and effective coupling strength. It can be seen that the entanglement decreases quickly as these parameters increase. Fig. \ref{Entanglement_upper_brunch}(b) shows how the logarithmic negativity increases with the cavity detuning frequency while it abruptly drops for large optical linewidth. The phenomenon remains consistent even though the change of cavity detuning frequency does not meet the parametric resonant condition $(\Delta = \Omega_M/2)$. This also justifies the fact the increase in the strength of parametric cavity drive $(E)$ reduces the cavity detuning and raises the squeezing parameter $(r)$. This in turn enhances the DEC-generated optomechanical coupling, which leads to a reduction of entanglement between cavity and mechanics. Finally, Fig. \ref{Entanglement_upper_brunch}(c) depicts the robustness of the entanglement against mechanical oscillation frequency and thermal noise. Therefore, the overall generated entangled in the upper bistability is three orders of magnitude greater than the one induced on the lower branch. However, this entanglement is somehow fragile, and this may be justified by decoherence coming from the nonlinear nature of the enhancement process.   Moreover, this entanglement in the upper branch transfers the squeezing of cavity mode to the mechanical oscillator.
	To get insight of the, we define a generalized quadrature of mechanics as  $ Q_M^\theta = ( e^{-i\theta} b +  e^{i\theta} b^\dagger )/\sqrt{2} $, and we calculate the fluctuation of the mechanical motion $(S_M^\theta = 1/2\langle \{Q_M^\theta, Q_M^\theta\} \rangle)$ as shown in Fig. \ref{wigfn_together}. The plot exhibits a clear dependency of the squeezing of mechanical motion on the quadrature phase angle $\theta$. This phenomenon can be also understood by analyzing the Wigner quasi-probability distribution in the phase spaces of the mechanical quadratures $(W (\mathbf{\tilde{\beta}}) = 2W(\mathbf{u_m})$ where $\mathbf{\tilde{\beta}} = [\Re\{\beta\},\Im\{\beta\}]^T + \sqrt{2}\mathbf{u_m}$ and  $\mathbf{u_M} = [Q, P]^T )$. The Brownian noise acting on the mechanical oscillator is considered to be a zero-mean Gaussian quantum stochastic process. Therefore, as a consequence, the linearized dynamics make the steady state of the system a zero-mean bipartite Gaussian state. The single mode correlation function which characterizes the motion of mechanics $(\mathbf{V_M} = \langle \{ \mathbf{u_M}, \mathbf{u_M}^T\}\rangle/2 )$, determines the Wigner function as,
	\begin{equation}
		W(\mathbf{u_m}) = \frac{1}{2\pi \sqrt{\det[\mathbf{V_M}]}} \exp \left[ -\frac{1}{2} \mathbf{u_M}^T \mathbf{V_M}^{-1} \mathbf{u_M}\right].
	\end{equation}
	
	The white dashed lines indicate the boundary of the Wigner function where the amplitude becomes half of the maximum possible amplitude, and the black dashed line corresponds to that for an ideal non-squeezed coherent state, and therefore, representing SQL. The fact that the boundary of the state, presented by white dashed lines going inside the boundary of the black one ensures the state to be squeezed below SQL. Fig. \ref{wigfn_together}(a) shows that the quadrature of squeezing is itself dependent on the mechanical drive, which also neutralizes while increasing. The sharp squeezed quadrature observed in the lower drive flattens while increasing its strength. Besides, the fluctuation in the quadrature of maximum variance reduces significantly in higher mechanical drive, bringing up a stable squeezing. This is also hinted from the plots of Wigner functions, where we see the direction of the squeezed quadrature rotates in phase space. A similar phenomenon is observed for the variation of DCE-generated optomechanical coupling in Fig. \ref{wigfn_together}(b).  This happens due to a better coherent transfer of cavity squeezing to mechanical mode with the increment of mechanical drive and optomechanical coupling. The enhancement of the influence of cavity squeezed mode on the mechanical oscillator saturates with further incrementation. The phenomenon is confirmed in Appendix \ref{Anal_num_compare} where we determined the quadrature fluctuations in two different ways and compared them. However, an opposite circumstance is experienced while increasing cavity detuning in Fig. \ref{wigfn_together}(c), as the squeezing sharpen, which is also witnessed from the plots of corresponding Wigner functions. This also concludes that both an increase of mechanical and cavity drives, have a similar impact on the dynamics. As a stronger cavity drive reduces cavity detuning and enhances cavity squeezing which increases effective optomechanical coupling, it causes strengthening mechanical squeezing.
	Furthermore, Fig. \ref{wigfn_together}(d) shows that the squeezing quadrature changes steadily with linewidth broadening. Therefore, the Wigner function shows the squeezed quadrature rotates in phase space, with a minimum change in squeezing amplitude. Opposite to cavity detuning, Fig. \ref{wigfn_together}(e) shows that the squeezing weakens with the increment of the mechanical oscillation frequency. Furthermore,  Fig. \ref{wigfn_together}(f) exhibits an increment of fluctuations in mechanical quadratures due to the injection of more thermal noise into the mechanical oscillator. The Wigner phase space boundary is finally seen to exceed SQL limits for higher thermal noise. Note, importantly, that the system parameters are chosen to be in the experimentally executable range which can be applicable in realistic optomechanical setup. As the parameters have been used before in \cite{DCE_optomechanics_Nori} and they satisfy the system’s stability condition for the feasible input cavity and mechanical drives, our results can then be implemented.
	
	\section{Conclussion}
	We investigated the possibilities of inducing the squeezing of mechanical motion and generating entanglement based on bistability in optomechanics. These nonclassical states have been engineered by using optical steady states lying on the lower and upper branches of the bistable shape. We found that the generated nonclassical states are weak when the optical steady states are on the lower branch.  By using the steady states on the upper branch, we observed a giant entanglement and a large squeezing going beyond the SQL.  We observe that the optomechanical entanglement decreases while increasing optomechanical squeezing and mechanical drive due to mechanical decoherence. The overall entanglement seems to be fragile due to the nonlinear effect involved in its enhancement process. The squeezing depends on the phase quadrature and shows robustness against thermal fluctuations. This work opens up a new way to generate nonclassical states by exploring nonlinear effects in optomechanics. This scheme can be also extended to similar fields such as in superconducting microwave and hybrid optomechanical systems.
	
	\begin{acknowledgments}
		S. Agasti wishes to acknowledge the European Union; Project number: 101065991 (acronym: SingletSQL) for supporting the work.
		
		P. Djorw\'e acknowledges the receipt of a grant from the APS-EPS-FECS-ICTP Travel Award Fellowship Programme (ATAP), Trieste, Italy.
	\end{acknowledgments}
	
	\appendix
	\begin{figure*}[t!]
		\includegraphics[width= 0.9 \linewidth]{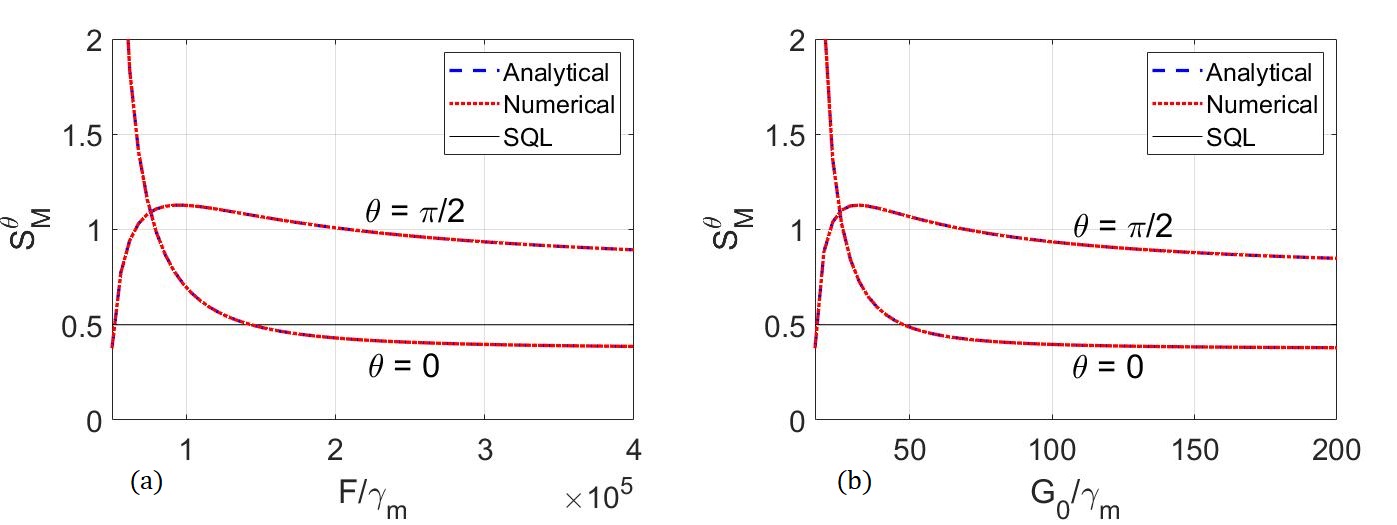}
		\caption{ Analytically and numerically determined squeezing with the variation of (a) mechanical drive and (b) effective optomechanical coupling.  
			The rest of all other parameters remain the same with Fig. \ref{cav_population_bistable} }
		\label{Anal_vs_num_nm=0}
	\end{figure*}
	
	\section{Appendix: Numerical vs Analytical estimation of mechanical squeezing} \label{Anal_num_compare}
	In the presence of a strong coherent tone, the dynamics of the system are affected by the fluctuations introduced by the nonlinear Hamiltonian. We determine the fluctuation spectra analytically from the linearized Heisenberg EOM over nonlinear stationary field amplitude. 
	
	\begin{equation}
		\label{eq:S11}
		\tilde{S}_M^\theta(\omega) = \frac{1}{2} \langle \left\{Q_M^\theta (\omega) , Q_M^\theta (-\omega)  \right\} \rangle ,
	\end{equation}
	with $ \tilde{Q}_M^\theta (\omega) = 1/\sqrt{2} \left( \tilde{b}_{-\omega}^\dagger e^{i \theta} + \tilde{b}_\omega e^{-i \theta}\right)$,
	with the usual definition of the Fourier transform:
	\begin{align*}
		\tilde{b}_\omega = \int_{-\infty}^\infty \mathrm{d}t e^{i\omega t} \tilde{b}(t), \qquad \qquad \tilde{b}_{-\omega}^{\dagger} = \int_{-\infty}^\infty \mathrm{d}t e^{i\omega t}\tilde{b}^{\dagger}(t)
	\end{align*}
	
	Similarly, the Fourier transform of cavity modes can also be defined in this way. Therefore, the equation of motion given in Eq. \eqref{EOM_t}, can be written in the Fourier domain as
	
	\begin{equation}\label{EOM_w}
		(i\omega +\mathbf{A_M}) \tilde{\mathbf{u}} = \tilde{\mathbf{n}}(\omega)
	\end{equation}
	which determines the spectra of the correlation matrix
	
	\begin{equation}
		\tilde{\mathbf{V}}(\omega) = \int \mathrm{d}\omega' \, \mathbf{\tilde{M}} (\omega) \mathbf{\tilde{D}}(\omega, \omega') \mathbf{\tilde{M}}^\dagger(\omega)
	\end{equation}
	where
	\begin{subequations}
		\begin{align}
			\mathbf{\tilde{M}}(\omega) &= 	(i\omega +\mathbf{A_M})^{-1} \qquad \qquad \text{and} \\
			\mathbf{\tilde{D}}(\omega,\omega') &=  \mathbf{D} \delta(\omega - \omega') 
		\end{align}
	\end{subequations}
	
	The variance of a mechanical quadrature can be determined by integrating over the fluctuation spectra: $ S_M^\theta  = \frac{1}{2\pi}\int_{-\infty}^\infty \mathrm{d}\omega \,\, \tilde{S}_M^\theta(\omega)$. A comparison between analytically and numerically determined quadrature variance is presented in Fig. \ref{Anal_vs_num_nm=0} where we see good agreement of both plots.
	The squeezing is seen to be shifted from phase to amplitude quadrature while increasing both the mechanical drive and effective optomechanical coupling.
	This also shows stability in squeezing in higher mechanical drive and optomechanical coupling	which happens from the fact that higher optomechanical coupling or mechanical drive enables a stable coherent transfer of cavity optomechanical squeezed mode to mechanics.

	\nocite{*}	
	\bibliography{Dynamical_Casimir_mechanical_squeezing}
	\bibliographystyle{apsrev4-2}
	
	\appendix

\end{document}